\documentclass[aps,graphicx,10pt,showkeys,showpacs,twocolumn]{revtex4}
\usepackage{amsmath}
\usepackage{amscd}
\usepackage{graphicx}
\usepackage{subfigure}
\usepackage{appendix}
\usepackage{multirow}
\usepackage{mathrsfs}

\begin{document}

\title[Short Title]{Shortcuts to adiabatic state transfer in time-modulated two-level non-Hermitian systems}

\author{Qi-Cheng Wu$^{1,\footnote{These authors contributed equally to this work.}}$}
\author{Jun-Long Zhao$^{1,*}$}
\author{Yan-Hui Zhou$^{1}$}
\author{Biao-Liang Ye$^{1}$}
\author{Yu-Liang Fang$^{1}$}
\author{Yi-Hao Kang$^{2}$}
\author{Qi-Ping Su$^{2}$}
\author{Zheng-Wei Zhou$^{3,4,5,}$\footnote{E-mail:  zwzhou@ustc.edu.cn}}
\author{Chui-Ping Yang$^{2,}$\footnote{E-mail: yangcp@hznu.edu.cn}}

\affiliation{
$^{1}$Quantum Information Research Center and Jiangxi Province Key Laboratory of Applied Optical Technology, Shangrao Normal University, Shangrao 334001, China\\
$^{2}$School of Physics, Hangzhou Normal University, Hangzhou\\
$^{3}$CAS Key Lab of Quantum Information, University of Science and Technology of China, Hefei 230026, China\\
$^{4}$Anhui Center for fundamental sciences in theoretical physics, University of Science and Technology of China, Hefei 230026, China\\
$^{5}$Hefei National Laboratory, University of Science and
Technology of China, Hefei 230088, China Zhejiang 311121, China}

\begin{abstract}
Nontrivial spectral properties of non-Hermitian systems can give
rise to intriguing effects that lack counterparts in Hermitian
systems. For instance, when dynamically varying system parameters
along a path enclosing an exceptional point (EP), chiral mode
conversion occurs. A recent study [Phys. Rev. Lett. 133, 113802
(2024)] demonstrates the achievability of pure adiabatic state
transfer by specifically selecting a trajectory in the system
parameter space where the corresponding evolution operator
exhibits a real spectrum while winding around an EP. However, the
intended adiabatic state transfer becomes fragile when taking into
account the effect of the nonadiabatic transition. In this work,
we propose a scheme for achieving robust and rapid adiabatic state
transfer in time-modulated two-level non-Hermitian systems by
appropriately modulating the system Hamiltonian and time-evolution
trajectory. Numerical simulations confirm that a complete
adiabatic transfer can always be achieved even under nonadiabatic
conditions after one period for different initialized adiabatic
states, and the scheme remains insensitive to moderate
fluctuations in control parameters. Therefore, this scheme offers
alternative approaches for quantum-state engineering in
non-Hermitian systems.
\end{abstract}


\maketitle

\section{Introduction}

In recent decades, there has been a growing interest in physical
systems described by non-Hermitian (NH) Hamiltonians due to their
unique spectral properties~\cite{NH1,NH2,NH3}. These systems can
exhibit points of extreme degeneracy in the spectrum, known as
exceptional points (EPs), where both eigenvalues and corresponding
eigenmodes
coalesce~\cite{EPs1,EPs2,EPs3,EPs4,vibrational-cooling}.
Structures supporting EPs have inspired numerous counterintuitive
phenomena such as extreme sensitivity to
perturbations~\cite{sensitivity}, loss-induced transparency
effects~\cite{lossinduced,invisibility}, special topological
structures~\cite{topology1,topology-encircling}, and so on.

Recently studies have also revealed that the variation of system
parameters along a path encircling EPs can induce chiral mode
switching~\cite{topology-encircling,Berry-encircling,Xu-encircling,Li-encircling,Arkhipov-encircling1,Ergoktas-encircling,Tang-encircling,
Hassan-encircling,Hassan-encircling2,Zhang-encircling,Nasari-encircling,Feilhauer-encircling}.
For instance, based on hopping, Li et al. demonstrated a robust
switching of chiral modes in coupled waveguides on a standard
silicon-on-insulator platform~\cite{Li-encircling}. Additionally,
Feilhauer et al. theoretically and experimentally demonstrated
that a chiral state transfer can be achieved by manipulation of a
dissipative Hamiltonian encircling an exceptional point
(EP)~\cite{Feilhauer-encircling}. Arkhipov et al. further
showcased the recovery of state flip symmetry in dissipative
systems by applying nonadiabatic transformations in multimode
systems and exploiting the spectral topology of hybrid
diabolic-exceptional points~\cite{Arkhipov-encircling1}. Moreover,
several studies have indicated that dynamical flip-state asymmetry
can occur even when dynamically approaching EPs without
necessarily encircling
them~\cite{Hassan-encircling,Hassan-encircling2,Nasari-encircling}.
Hassan et al. provided evidence that a sufficiently slow variation
of parameters away from an exceptional point can still result in
robust asymmetric state exchange~\cite{Hassan-encircling}. More
recently,  in contrast to chiral or asymmetric mode conversion
schemes, Arkhipov et al. achieved symmetric state transfer by
intentionally selecting a trajectory in the system parameter space
where the corresponding evolution operator acquires a real
spectrum~\cite{Arkhipov-encircling2}. The aforementioned
conventional encircling schemes offer promising pathways for
manipulating quantum states in NH domains. However, a common
challenge encountered with these schemes is the difficulty in
achieving perfect adiabatic evolution in certain experiments,
where the slow nature of adiabatic evolution may lead to
decoherence effects due to interactions between the quantum system
and its environment~\cite{open1,open2,open3}. Therefore, it is
crucial to explore innovative approaches that demonstrate both
robustness and efficiency, serving as viable alternatives or
enhancements to conventional encircling techniques.

The previously-reported generalized approximate adiabaticity
criterion in the NH system can be expressed
as~\cite{criterion1,criterion2,criterion3,criterion4,criterion5}
\begin{eqnarray}\label{eq0-1}
\sum_{n\neq{m}}\frac{|{\langle\widehat{{\phi_{n}}}(t)|\dot{\phi}_{m}(t)\rangle}|}
{|\omega_{nm}(t)|}e^{i\int_{0}^{t}{\omega_{nm}}(\tau)d{\tau}}\ll
1,
\end{eqnarray}
where $\{\langle\widehat{{\phi_{n}}}(t)|\}$ and
$\{|{\phi}_{m}(t)\rangle \}$ are left and right eigenstates of the
NH Hamiltonian $H(t)$, $\omega_{nm}(t)$=$E_{{m}}(t)$-$E_{{n}}(t)$
is the difference between corresponding eigenvalues of the NH
Hamiltonian, and
${{\langle\widehat{{\phi_{n}}}(t)|\dot{\phi}_{m}(t)\rangle}}$ is
associated with the so-called nonadiabatic coupling transition
between the instantaneous eigenstates.  It is worth noting that in
NH systems, the eigenvalues are typically complex values. The
exponential function of the difference in imaginary parts of the
eigenvalues ($\textrm{exp}[-\textrm{Im}[{\omega_{nm}}(t)]]$) can
lead to either damping or amplification of the evolving state.
Moreover, even a trivial difference between two eigenvalues can
result in nontrivial changes in system dynamics if the trajectory
of adiabatic evolution crosses EPs. Generally, nonadiabatic
couplings become more pronounced as the speed of system evolution
increases without external control. Interestingly, these
nonadiabatic couplings can be nullified by applying so-called
shortcuts to adiabaticity (STA)~\cite{STA1,STA2,STA3}. The
critical idea behind STA is to accelerate quantum system dynamics
through a designed coherent control, such that the system evolves
on timescales much shorter than decoherence times. Based on this
innovative concept, several methods have been
proposed~\cite{TQD1,TQD2,TQD3,TQD4,LR1,LR2,LR3,Fast1,Fast2},
including counter-diabatic driving (equivalently known as
transitionless quantum algorithm)~\cite{TQD1,TQD2,TQD3,TQD4},
Lewis-Riesenfeld inverse engineering~\cite{LR1,LR2,LR3}, and
"fast-forward" scaling techniques~\cite{Fast1,Fast2}, and so on.

Motivated by the remarkable advancements in state transfer within
non-Hermitian systems through dynamic encircling of exceptional
points~\cite{topology-encircling,Berry-encircling,Xu-encircling,Li-encircling,Arkhipov-encircling1,Ergoktas-encircling,Tang-encircling,
Hassan-encircling,Hassan-encircling2,Zhang-encircling,Nasari-encircling,Feilhauer-encircling}
and shortcuts to
adiabaticity~\cite{criterion1,criterion2,criterion3,criterion4,criterion5},
we aim to address the following question: Can perfect shortcuts to
adiabatic state transfer be achieved in time-modulated
non-Hermitian systems? The answer is affirmative. In this work, we
demonstrate how a modified time-modulated non-Hermitian system can
be used to realize shortcuts to adiabatic state transfer by
encircling approximate EPs. Specifically, the adiabatic states
correspond to eigenstates of a time-modulated two-level
non-Hermitian Hamiltonian $H_{0}$ with EPs, while the designed
modified time-modulated non-Hermitian Hamiltonian $H_{m}$ for
shortcuts to adiabacity does not possess any exact EP but exhibits
an approximate EP where the eigenenergy spectrum is minimally
separated. By designing a trajectory for the modified system's
time evolution that closely approaches this approximate EP, robust
and rapid adiabatic state transfer can be faithfully achieved.
Different from conventional encircling
schemes~\cite{topology-encircling,Berry-encircling,Xu-encircling,Li-encircling,Arkhipov-encircling1,Ergoktas-encircling,Tang-encircling,
Hassan-encircling,Hassan-encircling2,Zhang-encircling,Nasari-encircling,Arkhipov-encircling2,Feilhauer-encircling}
which work under the adiabatic approximation condition, our scheme
could nullify potential nonadiabatic coupling by applying some
designed coherent controls, then it works well under the
nonadiabatic approximation condition. A complete adiabatic
transfer can always be obtained after one period for different
parameters and initialized states, and the initial adiabatic state
can eventually return to itself after two periods. Moreover, the
scheme demonstrates a good control performance and robustness in
view of fluctuations of control parameters.

The rest sections of this paper are organized as follows. In
Section~\ref{section:II}, we provide a concise overview of a
symmetric state transfer scheme in a two-level NH Hamiltonian
system, accompanied by an in-depth analysis of the effects of
non-adiabatic transitions. Section~\ref{section:III} is dedicated
to explicit discussions on how to engineer the NH Hamiltonian for
constructing shortcuts to adiabatic state transfer and designing
appropriate trajectories for system time evolution away from
approximate exceptional points (EPs). The feasibility and
performance of shortcuts to adiabatic state transfer are
comprehensively discussed step by step. Finally, we present a
summary in Section~\ref{section:IV}.

\section{Conventional adiabatic state transfer}\label{section:II}

Consider a two-level NH Hamiltonian~\cite{Arkhipov-encircling2}
\begin{eqnarray}\label{eq1-1}
H_{0}=(k+i\kappa)\sigma_{x}+(\varepsilon-i \Delta)\sigma_{z},
\end{eqnarray}
where $\sigma_{x}$ and $\sigma_{z}$ are Pauli matrices, and
$k,\kappa,\varepsilon,\Delta \in \mathbf{R}$.  A possible
realization of such a system is two coupled dissipative cavities
(in the mode
representation)~\cite{Arkhipov-encircling2,experiment-realize1,experiment-realize2,experiment-realize3},
where $\Delta$ $(-\Delta)$ denotes the resonator gain (loss) rate
and $\varepsilon$ is the frequency detuning of the resonators. The
resonators are coupled coherently with interaction strength $k$,
while $\kappa$  accounts for dissipative coupling strength.

The eigenvalues of the NH Hamiltonian
\begin{eqnarray}\label{eq1-2}
E_{\pm}&=&\pm\sqrt{E_1 E_2},\cr
E_1&=&k(t)-\Delta-i(\varepsilon-\kappa),\cr
E_2&=&k(t)+\Delta+i(\varepsilon+\kappa),
\end{eqnarray}
are complex, and one can not intuitively obtain much information
about them except EPs $\{\Delta_{EP}$=$k,
\varepsilon_{EP}$=$\kappa\}$ and $\{\Delta_{EP}$=$-k,
\varepsilon_{EP}$=$-\kappa\}$. However, with the help of a certain
function $f: \vec{r}=(x,y)\rightarrow
(k,\kappa,\varepsilon,\Delta)$ as
follows~\cite{Arkhipov-encircling2,map1,map2}:
\begin{eqnarray}\label{eq1-3}
k=\alpha\cosh\phi_{i}\sin\phi_{r},
\kappa=\alpha\sinh\phi_{i}\cos\phi_{r},\cr
\varepsilon=\alpha\cosh\phi_{i}\cos\phi_{r},
\Delta=\alpha\sinh\phi_{i}\sin\phi_{r},
\end{eqnarray}
where $\phi=\phi_{r}+i\phi_{i}=\arctan[(x+iy)^{-1}]\in
\mathbf{C}$, $\alpha=x\sinh\phi_{i}/\sin\phi_{r}\in \mathbf{R}$,
the NH Hamiltonian in Eq.~(\ref{eq1-1})  acquires a simplified
form
\begin{eqnarray}\label{eq-H0}
H_{0}= \alpha(t)\left(\begin{array}{ccccccc}
\cos{\phi} &  \sin{\phi}\\
\sin{\phi} & -\cos{\phi} \\
\end{array}\right).
\end{eqnarray}
It is not hard to find that $H_{0}$ has real eigenvalues
$E_{\mp}=\mp\alpha$ and the corresponding right eigenvectors  are
\begin{eqnarray}\label{eq1-5}
|\phi_{-}\rangle&=&[-\sin\frac{\phi}{2},\cos\frac{\phi}{2}]^{T},\cr
|\phi_{+}\rangle&=&[\cos\frac{\phi}{2},\sin\frac{\phi}{2}]^{T},
\end{eqnarray}
where \textit{T} stands for transpose. Together with the left
eigenvectors
\begin{eqnarray}\label{eq1-6}
|\widehat{\phi_{-}}\rangle&=&[-\sin\frac{\phi^{*}}{2},\cos\frac{\phi^{*}}{2}]^{T},\cr
|\widehat{\phi_{+}}\rangle&=&[\cos\frac{\phi^{*}}{2},\sin\frac{\phi^{*}}{2}]^{T},
\end{eqnarray}
where the asterisk means complex conjugate. The biorthogonal
partners $\{\langle\widehat{{\phi_{n}}}|\}$ and
$\{|{\phi}_{m}\rangle \}$ $(n,m$=$+,-)$ are normalized to satisfy
the biorthogonality relation~\cite{NH3,Biorthogonal1}
\begin{eqnarray}\label{eq1-7}
\langle\widehat{{\phi_{n}}}|\phi_{m}\rangle=\delta_{nm},
\end{eqnarray}
and the closure relation
\begin{eqnarray}\label{eq1-8}
\sum_{n}|\widehat{{\phi_{n}}}\rangle\langle\phi_{n}|=\sum_{n}|{\phi_{n}}\rangle\langle\widehat{{\phi_{n}}}|=1.
\end{eqnarray}
\subsection{System eigenenergy spectrum and time-evolution trajectory}
\begin{figure}
\scalebox{0.44}{\includegraphics{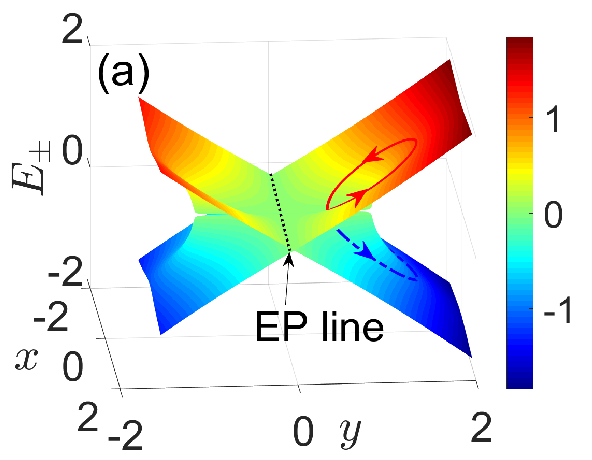}\includegraphics{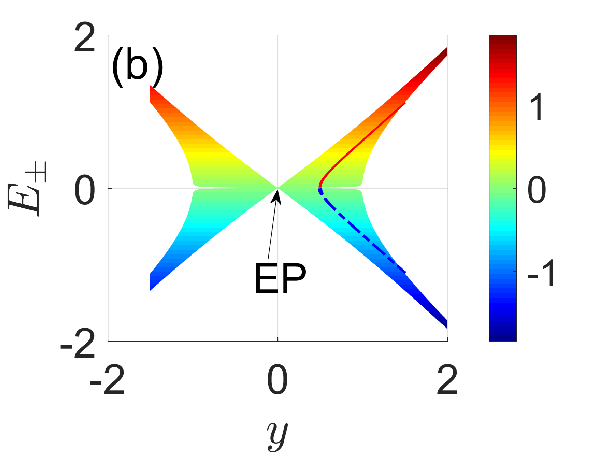}}
\scalebox{0.44}{\includegraphics{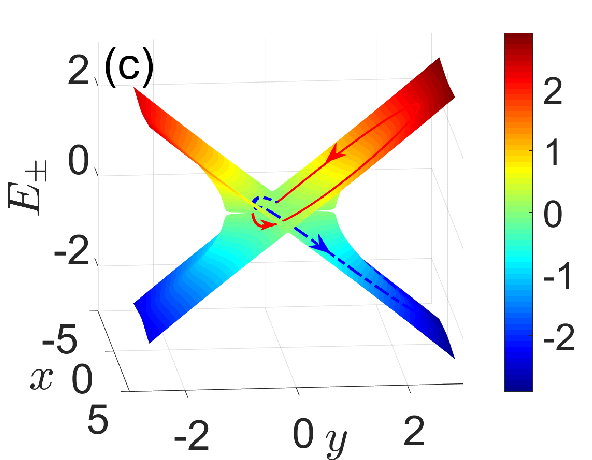}\includegraphics{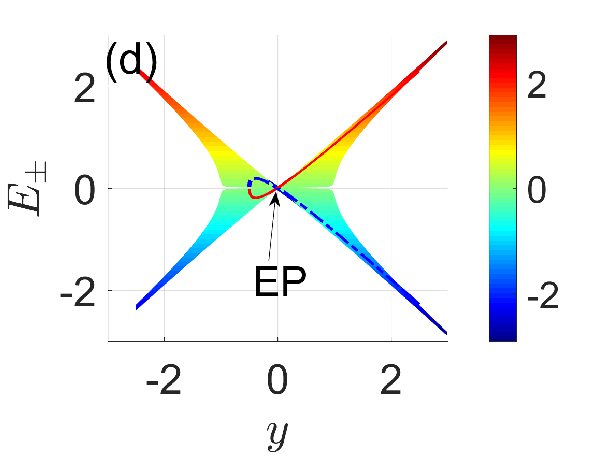}}
\caption{\label{fig-H0-eigenenergyspectrum} The eigenenergy
spectrum $E_{\pm}(x,y)$ and the system's time-evolution trajectory
$E_{\pm}(x(t),y(t))$ of the non-Hermitian Hamiltonian $H_{0}(t)$
given in Eq.~(\ref{eq-H0}) are depicted. The two energy Riemann
surfaces (the red surface corresponds to $E_{-}(x,y)$ and the blue
surface corresponds to $E_{+}(x,y)$) are connected by a line of
exceptional points at $y=0$. The solid red line represents
$E_{-}(t)$, while the dashed blue line denotes $E_{+}(t)$. In
panels (a) and (b) [(c) and (d)], identical parameters
($r=0.5,\omega=\pi/100,\phi_{0}=\pi$)[($r=1.5,\omega=\pi/100,\phi_{0}=\pi$)]
are used. The eigenenergy spectrum is presented as a
three-dimensional surface in panels (a) and (c), whereas it is
shown as a two-dimensional plot in panels (b) and (d).}
\end{figure}

To gain an intuitive understanding of the energy change in the
system, we present the eigenenergy spectrum $E_{\pm}$ in the
parameter space $(x, y)$ using both three-dimensional plots
[Figs.~\ref{fig-H0-eigenenergyspectrum}(a) and (c)] and
two-dimensional plots [Figs.~\ref{fig-H0-eigenenergyspectrum}(b)
and (d)]. As depicted in Figs.~\ref{fig-H0-eigenenergyspectrum}(a)
and (c), two energy Riemann surfaces are connected by an
exceptional line $y=0$, where all exceptional points reside.
Mathematically, the phase parameter
$\phi=\phi_{r}+i\phi_{i}=\arctan[1/x]$ is a real parameter when
setting $y=0$. Consequently, it follows that $\sinh\phi_{i}=0$ and
$\alpha=x\sinh\phi_{i}/\sin\phi_{r}=0$. This observation is
remarkable as it implies that EPs exist extensively in the
parameter space $(x, y)$, allowing for easy identification of
appropriate system time-evolution trajectories that encircle or
approach these EPs.

For instance, a system time-evolution trajectory can be chosen
as~\cite{Arkhipov-encircling2}
\begin{eqnarray}\label{eq1-9}
x(t)=r\sin(\omega{t}+\phi_{0}),y(t)=1-r\cos(\omega{t}+\phi_{0}),
\end{eqnarray}
where $r,w, \phi_{0} \in \mathbf{R}$ are constants, and the time
$t$ is presented in arbitrary units. Based on Eq.~(\ref{eq1-9}),
the NH Hamiltonian $H_{0}(t)$ undergoes periodic changes with
respect to time $t$ (with a period of $T$=$2\pi/\omega$). The
time-evolution trajectories of $E_{\pm}(t)$ are also depicted in
Fig.~\ref{fig-H0-eigenenergyspectrum}, where the solid red line
represents $E_{-}(t)$ and the dashed blue line denotes $E_{+}(t)$.
The initial phase $\phi_{0}$=$\pi$ corresponds to the maximum
separation between the two energy levels, $E_{\pm}(t)$. It can be
observed that both orbiting trajectories of $E_{\pm}(t)$ form
irregular circles around the EPs, undergoing slight shape
variations near these points. For positive angular frequencies
($\omega>0$), the orbiting trajectory proceeds counterclockwise
(or clockwise). Furthermore, we emphasize that the radius, denoted
as $r$, plays a crucial role in this dynamic process. When $r<1$
[see Figs.~\ref{fig-H0-eigenenergyspectrum} (a) and (b)], the
loops of time-evolution trajectory for $E_{\pm}(t)$ remain distant
from the exceptional lines, moving solely within a fixed plane of
energy spectrum. However, when $r \geq 1$ [see
Fig.~\ref{fig-H0-eigenenergyspectrum} (c) and (d)], these loops
cross over at $(\pm \sqrt{r^2-1}, 0)$ onto an opposite plane of
energy spectrum during $\tau\subset[\tau_1|x(\tau_1)
=-\sqrt{r^2-1}, \tau_2|x(\tau_21)=\sqrt{r^2-1}]$, before returning
back to their original plane in energy spectrum. It is worth
noting that there is only a small difference in eigenvalues
$\omega_{nm}(t)$ when approaching times close to $\tau_1,\tau_2$.

\subsection{Performance of adiabatic state transfer}

\begin{figure}
\scalebox{0.44}{\includegraphics{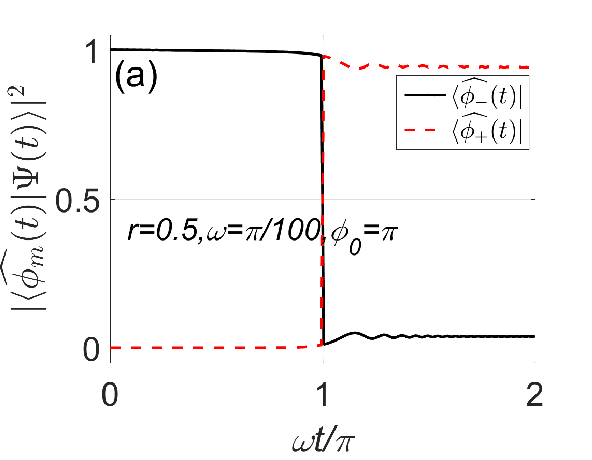}\includegraphics{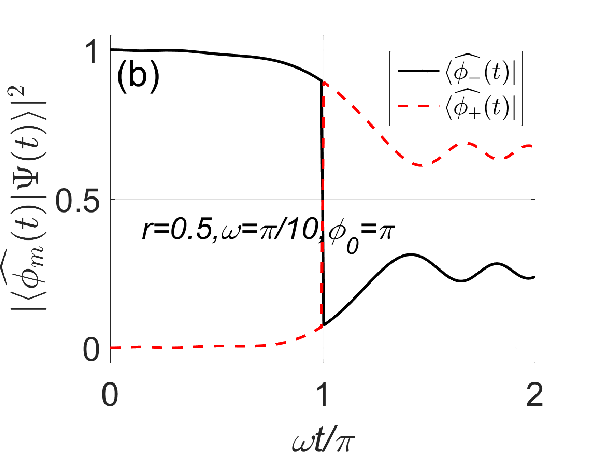}}
\scalebox{0.44}{\includegraphics{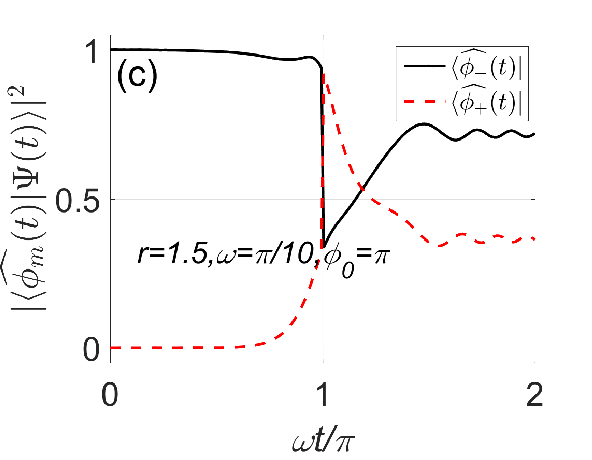}\includegraphics{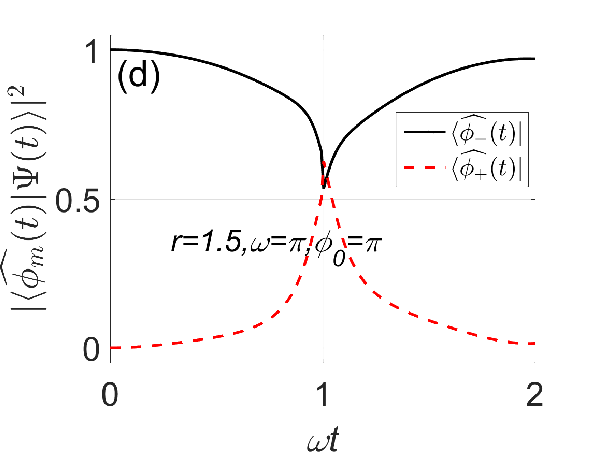}}
\caption{\label{fig-adiabaticstatetransfer} The time evolution of
fidelity $F_{n}=|\langle\widehat{{\phi_{n}}}(t)|\Psi(t)\rangle|^2$
for the time-depending right eigenstate $|\phi_{n}(t)\rangle$
$(n$=$+,-)$. The initial state is chosen as
$|\Psi(0)\rangle$=$|\phi_{-}(0)\rangle$ [see Eq.~(\ref{eq1-5})]
and other parameters are set as follows: (a)$\{r=0.5$,
$\omega=\pi/100$, $\phi_{0}=\pi\}$, (b)$\{r=0.5$, $\omega=\pi/10$,
$\phi_{0}=\pi\}$,(c)$\{r=1.5$, $\omega=\pi/10$, $\phi_{0}=\pi\}$,
(d)$\{r=1.5$, $\omega=\pi$, $\phi_{0}=\pi\}$.}
\end{figure}

Now we proceed to elaborate on the performance of the adiabatic
state transfer while dynamically encircling an exceptional point.
The fidelity for the right eigenstate $|\phi_{n}(t)\rangle$
$(n$=$+,-)$ is determined by the relation
$F_{n}$=$|\langle\widehat{{\phi_{n}}}(t)|\Psi(t)\rangle|^2$, where
$|\Psi(t)\rangle$ represents the evolving state of the system at
time $t$. Assuming that the system is initially prepared in one of
its right eigenstates, i.e.,
$|\Psi(t$=$0)\rangle$=$|\phi_{-}(t$=$0)\rangle$, we can obtain the
evolving state $|\Psi(t)\rangle$ through numerical integration of
Schr\"{o}dinger's equation. The instantaneous left eigenvector
$\langle\widehat{{\phi_{n}}}(t)|$ can be calculated by
substituting the parameters into Eqs.~(\ref{eq1-6}) and
(\ref{eq1-9}).

For the sake of convenience, in
Fig.~\ref{fig-adiabaticstatetransfer}, we present the temporal
evolution of the fidelity for eigenstate states with four sets of
parameters: (a)$\{r=0.5$, $\omega=\pi/100$, $\phi_{0}=\pi\}$,
(b)$\{r=0.5$, $\omega=\pi/10$, $\phi_{0}=\pi\}$,(c)$\{r=1.5$,
$\omega=\pi/10$, $\phi_{0}=\pi\}$, (d)$\{r=1.5$, $\omega=\pi$,
$\phi_{0}=\pi\}$. Figure~\ref{fig-adiabaticstatetransfer}(a)
exhibits a relatively complete eigenstate transfer, where
$|\phi_{-}(t)\rangle$$\leftrightarrow$$|\phi_{+}(t)\rangle$ are
exchanged after one period $T$=$2\pi/\omega$. However, it is
regrettable that the final fidelity falls short for practical
applications. Moreover, as depicted in
Figs.~\ref{fig-adiabaticstatetransfer}(b)-(d), the state transfer
fails entirely when there are dynamic variations in the parameters
$(x(t),y(t))$, i.e., when $\omega$ or $r$ takes on large values.
In fact, there are two potential physical mechanisms involved in
the entire dynamic process: (i) under perfect adiabatic
conditions, if the system's time-evolution trajectory winds around
or approaches EPs, a perfect state transformation occurs within
half a period, and the transferred state will be preserved due to
the adiabatic effect. (ii) When the adiabatic conditions are not
perfectly satisfied, nonadiabatic coupling transitions between
instantaneous eigenstates always occur, leading to the destruction
of desired state transfer throughout the evolutionary process. It
is worth noting that the adiabatic conditions [see
Eq.~(\ref{eq0-1})] is also associated with the difference in
eigenvalues $\omega_{nm}(t)$ of the system. If the system's
time-evolution trajectory passes through EPs
($x(t)=\pm\sqrt{r^2-1},y(t)=0$), as shown in
Fig.~\ref{fig-H0-eigenenergyspectrum} (c) and (d), temporarily
amplified effects of nonadiabatic coupling transitions may arise.

The aforementioned mechanisms can be employed to comprehend the
phenomena depicted in Fig.~\ref{fig-adiabaticstatetransfer}. In
Fig.~\ref{fig-adiabaticstatetransfer}(a), under a suitable
adiabatic condition, mechanism (i) predominantly governs the
system evolution, while the adverse impact of mechanism (ii) leads
to fidelity fluctuations throughout the process. Furthermore, upon
comparing Fig.~\ref{fig-adiabaticstatetransfer}(a) with
Fig.~\ref{fig-adiabaticstatetransfer}(b), it becomes evident that
the adiabatic conditions in
Fig.~\ref{fig-adiabaticstatetransfer}(b) are worse due to a large
$\omega$, thereby amplifying the role of mechanism (ii) during the
system evolution and causing rapid fidelity degradation. Moreover,
in Fig.~\ref{fig-adiabaticstatetransfer}(c), when $r$=1.5$>$1, it
implies that the time-evolution trajectory crosses the exceptional
line at $(\pm\sqrt{r^2-1},0)$, triggering two instances of
significant added-state transformations through enhanced
manifestation of mechanism (ii). Consequently, compared to results
obtained from Fig.~\ref{fig-adiabaticstatetransfer}(b),
significant state transitions occur precisely when
$\tau_{1}|x(\tau_{1})=-\sqrt{r^2-1}$ and
$\tau_{2}|x(\tau_{21})=\sqrt{r^2-1}$. Finally, in
Fig.~\ref{fig-adiabaticstatetransfer}(d), due to an extremely poor
performance of adiabetic conditions, mechanism (ii) overwhelmingly
dominates the system evolution. Therefore, achieving flawless
state transition necessitates minimizing the nonadiabetic coupling
transition effects as much as possible and avoiding the trajectory
passage through exceptional points during the system evolution.

\section{Shortcuts to adiabatic state transfer}\label{section:III}

It is advisable to project the system into the so-called adiabatic
frame $\{|\phi_{+}(t)\rangle,|\phi_{-}(t)\rangle\}$, with
$H^{a}_{0}(t)=\tilde{R}^{\dag}(t)H_{0}(t)R(t)-i\hbar
\tilde{R}^{\dag}(t)\dot{R}(t)$, where the transformation matrixes
$R(t)$ and $\tilde{R}^{\dag}(t)$ are given by
\begin{eqnarray}\label{eq3-1}
R(t)=\left(\begin{array}{ccccccc}
\cos\frac{\phi(t)}{2} &  -\sin\frac{\phi(t)}{2}\\
\sin\frac{\phi(t)}{2} & \cos\frac{\phi(t)}{2}  \\
\end{array}\right),\cr \tilde{R}^{\dag}(t)
=\left(\begin{array}{ccccccc}
\cos\frac{\phi(t)}{2} &  \sin\frac{\phi(t)}{2}\\
-\sin\frac{\phi(t)}{2} & \cos\frac{\phi(t)}{2}  \\
\end{array}\right).
\end{eqnarray}
Then, the original Hamiltonian $H_{0}(t)$ in the adiabatic basis
can be written as
\begin{eqnarray}\label{eq3-2}
H^{a}_{0}(t)=\left(\begin{array}{ccccccc}
E_{+}(t) &  0\\
0 & E_{-}(t)  \\
\end{array}\right)-i\hbar\left(\begin{array}{ccccccc}
0 &  -\frac{\dot{\phi}(t)}{2}\\
\frac{\dot{\phi}(t)}{2} & 0 \\
\end{array}\right),
\end{eqnarray}
where
\begin{eqnarray}\label{eq3-4}
\dot{\phi}(t)&=&-\frac{[\dot{x}(t)a(t)+\dot{y}(t)b(t)+i(\dot{y}(t)a(t)-\dot{x}(t)b(t))]}{a^2(t)+b^2(t)},\cr
a(t)&=&1+x^{2}(t)+y^{2}(t),~~ b(t)=2x(t)y(t).
\end{eqnarray}
The adiabatic process can be regarded as a two-level toy model
driven simultaneously by a coherent field $\Omega_{0}(t)$ and an
incoherent field $\Omega_{1}(t)$. It is evident that under the
conditions
\begin{eqnarray}\label{eq3-4}
|\Omega_{0}(t)|,|\Omega_{1}(t)|\ll E_{+}-E_{-},
\end{eqnarray}
with $\Omega_{0}(t)$=$\textrm{Re}[\dot{\phi}(t)/2]$ and
$\Omega_{1}(t)$=$\textrm{Im}[\dot{\phi}(t)/2]$, the transitions
between the adiabatic states will be significantly suppressed due
to rapid oscillations. However, in the event that the condition
fails to meet the required criteria, nonadiabatic coupling
transitions may emerge as the predominant factor influencing the
system evolution.

\subsection{Modified effective Hamiltonian and time-evolution trajectory}
The STA theory (transitionless quantum
driving~\cite{STA1,STA2,STA3,TQD1,TQD2,TQD3,TQD4}) provides an
effective approach to achieve the state transfer without relying
on the adiabatic condition through modifications to the system's
Hamiltonian.  Moreover, the modified effective Hamiltonian can be
calculated by
\begin{eqnarray}\label{eq3-5}
{H}_m(t)&=&{H}_{0}(t)+{H}_{1}(t),
\end{eqnarray}
where
\begin{eqnarray}\label{eq3-6}
{H}_1(t)&=&i\hbar \tilde{R}^{\dag}(t)\dot{R}(t)\cr
&=&i\hbar\left(\begin{array}{ccccccc}
0 &  -\frac{\dot{\phi}(t)}{2}\\
\frac{\dot{\phi}(t)}{2} & 0 \\
\end{array}\right).
\end{eqnarray}
Theoretically, the system is driven along adiabatic paths defined
by ${H}_{0}(t)$ through ${H}_m(t)$. However, implementing the
added term ${H}_1(t)$=$\dot{\phi}(t)/2\sigma_{y}$ in practice is
not straightforward. Specifically, $\dot{\phi}(t)$ is often a
complex number, resulting in non-complex conjugate off-diagonal
terms. Therefore, it is advisable to set
$\dot{\phi}(t)$=$\textrm{Re}[\dot{\phi}(t)]$, which offers two
distinct advantages: (i) the practical realization of ${H}_1(t)$
becomes easier (here ${H}_1(t)$ can be considered as a coherent
field $\Omega_{0}(t)$ with a fixed phase $e^{\frac{\pi}{2}}$);
(ii) the eigenvalues of the modified effective Hamiltonian
${H}_m(t)$ remain real values
$E'_{\pm}$=$\pm\sqrt{\textrm{Re}[\dot{\phi}(t)]^{2}/4+\alpha^2(t)}$
across all parameter ranges $(x(t),y(t))$.

The elimination of $\textrm{Im}[\dot{\phi}(t)]$ can be achieved by
satisfying the following equation
\begin{eqnarray}\label{eq3-7}
-\dot{x}(t)b(t)=\dot{y}(t)a(t)),
\end{eqnarray}
with $a(t)$=$1$+$x^{2}(t)$+$y^{2}(t)$, $b(t)$=$2x(t)y(t)$. After
performing algebraic calculations, we derive a modified trajectory
for the time evolution of the system
\begin{eqnarray}\label{eq-modifiedtrajectory}
x(t)=r\sin(\omega{t}+\phi_{0}),\cr
y'(t)=\sqrt{1+r^2}-r\cos(\omega{t}+\phi_{0}).
\end{eqnarray}
We emphasize that the modified trajectory for the time-evolution
of the system cloud ensure the elimination of
$\textrm{Im}[\dot{\phi}(t)]$ and prevents the system
time-evolution trajectory from crossing the exceptional line $y$=0
of $H_{0}(t)$ ($y'(t)$$\neq$0). It should be noted that the choice
of the trajectory in Eq.~(\ref{eq-modifiedtrajectory}) is not
unique, allowing for flexibility based on different situations.

Up to now, we have successfully accomplished the design of
shortcuts to adiabatic state transfer and system time-evolution
trajectory, yielding the subsequent modified NH Hamiltonian
\begin{eqnarray}\label{eq-Hm}
H_{m}(t)=[k(t)+i\kappa(t)]\sigma_{x}+\Omega(t)\sigma_{y}+[\varepsilon(t)-i
\Delta(t)]\sigma_{z},
\end{eqnarray}
where
\begin{eqnarray}\label{eq3-10}
k(t)&=&\alpha(t)\cosh\phi_{i}(t)\sin\phi_{r}(t),\cr
\kappa(t)&=&\alpha(t)\sinh\phi_{i}(t)\cos\phi_{r}(t),\cr
\varepsilon(t)&=&\alpha(t)\cosh\phi_{i}(t)\cos\phi_{r}(t),\cr
\Delta(t)&=&\alpha(t)\sinh\phi_{i}(t)\sin\phi_{r}(t),\cr
\Omega(t)&=&\dot{\phi}(t)/2,
\end{eqnarray}
with
$\phi(t)$=$\phi_{r}(t)$+$i\phi_{i}(t)$=$\arctan[(x(t)$+$iy'(t))^{-1}]$
$\in$ $\mathbf{C}$,
$\alpha(t)$=$x(t)\sinh\phi_{i}(t)/\sin\phi_{r}(t)$,$\dot{\phi}(t)$$\in$
$\mathbf{R}$. From an experimental perspective, the modified NH
Hamiltonian can be derived in the two-mode approximation for
counter-traveling waves in a whispering-gallery microcavity, such
as a micro-disk or micro-toroid, perturbed by $N$ Rayleigh
scatterers~\cite{experiment-realize1,experiment-realize2,experiment-realize3}.
The effective Hamiltonian in the traveling-wave basis
(counterclockwise (CCW), clockwise (CW)) is.
\begin{eqnarray}\label{eq3-11}
H^{(M)}=\left(\begin{array}{ccccccc}
\omega^{(M)} &  A^{(M)}\\
B^{(M)} & \omega^{(M)}  \\
\end{array}\right),
\end{eqnarray}
where $\omega^{(M)}$=$\omega_{0}$+$\sum^{M}_{j=1}\varepsilon_{j}$,
$A^{(M)}$=$\sum^{M}_{j=1}\varepsilon_{j}e^{-i2k\beta_{j}}$ and
$B^{(M)}$=$\sum^{M}_{j=1}\varepsilon_{j}e^{i2k\beta_{j}}$. Here,
 $m$ is the azimuthal mode number, $\omega_{0}$ is the complex
frequency of the unperturbed resonance mode, $\beta_{j}$ is the
angular position of scatterer $j$, and $\varepsilon_{j}$ is the
complex frequency splitting that is introduced by scatterer $j$
alone. Notably, due to the asymmetry in backscattering between
clockwise- and anticlockwise-travelling waves, it is possible for
$|A^{(M)}|$ to differ from $|B^{(M)}|$.

\subsection{Performance of shortcuts to adiabatic state transfer}

\begin{figure}
\scalebox{0.44}{\includegraphics{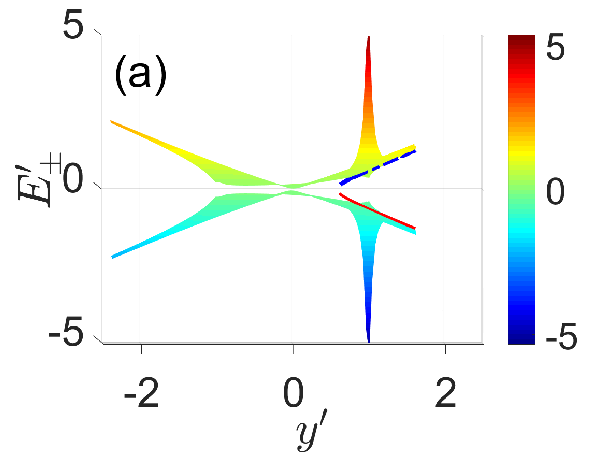}\includegraphics{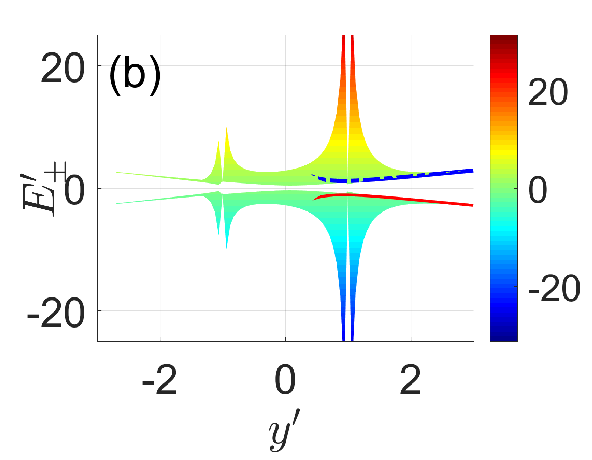}}
\caption{\label{fig-Hm-eigenenergyspectrum} The eigenenergy
spectrum $E'_{\pm}(x,y')$ and the system's time-evolution
trajectory $E'_{\pm}(x(t),y'(t))$ of the NH Hamiltonian $H_{0}(t)$
given in Eq.~(\ref{eq-H0}) are presented. Two energy surfaces,
with the blue surface corresponding to $E'_{-}(x,y)$ and the red
surface corresponding to $E'_{+}(x,y)$, are clearly separated. The
solid red line represents $E'_{-}(t)$ while the dashed blue line
denotes $E'_{+}(t)$. In (a) and (b), we consider parameters
($r=0.5,\omega=\pi/10,\phi_{0}=\pi$) and
($r=1.5,\omega=\pi/10,\phi_{0}=\pi$), respectively.}
\end{figure}

\begin{figure}
\scalebox{0.44}{\includegraphics{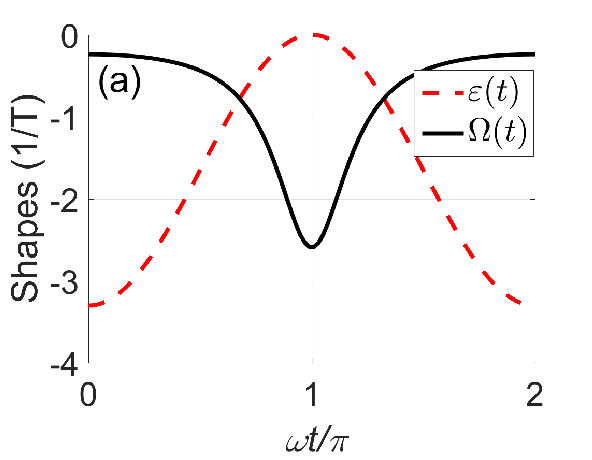}\includegraphics{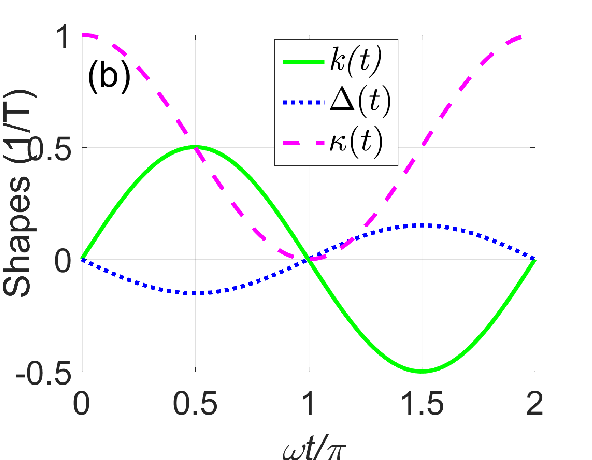}}
\caption{\label{Shape}  The shapes of the time-modulated system
parameters in Eq.~(\ref{eq3-10}) are depicted when winding
counterclockwise in the chart $(x, y')$, as described by
Eq.~(\ref{eq-modifiedtrajectory}). (a) The frequency detuning
$\varepsilon(t)$ is represented by a red dashed curve, while the
coupling strength $\Omega(t)$ is shown as a black solid curve. (b)
The coherent mode coupling strength $k(t)$ is illustrated with a
green solid curve, and the gain-loss rate $\Delta(t)$ is displayed
using a blue dot dash curve. Additionally, the incoherent mode
coupling strength $\kappa(t)$ is presented as a pink  dashed
curve. The parameters have been set to
${r=1.5,\omega=\pi/10,\phi_{0}=\pi}$.}
\end{figure}

Before elaborating on the performance of the scheme for a complete
fidelity transfer, we also briefly evaluate the eigenenergy
spectrum of ${H}_m(t)$. We plot the eigenenergy spectrum
$E'_{\pm}$=$\pm\sqrt{\Omega^{2}(t)+\alpha^{2}(t)]}$ in the
parameter space $(x, y')$. As shown in
Figs.~\ref{fig-Hm-eigenenergyspectrum}(a) and (b), two energy
surfaces (the red surface belonging to $E_{+}$ and blue surface
belonging to $E_{-}$) are separated. This indicates that there are
no EPs in the modified NH Hamiltonian. Although $y'$=0 is no
longer an EP line, the two eigenenergy spectra $E_{\pm}(x, y')$
remain minimally separated near it, which can be considered as an
approximate EP (AEP) line. Importantly, regardless of arbitrary
choices for radius $r$ and frequency $\omega$, the system's
time-evolution trajectory will not cross this AEP line. This
result has immediate implications for potentially reducing
experimental implementation difficulties associated with certain
parameters. The shapes of the time-modulated system parameters in
Eq.~(\ref{eq-Hm}) are plotted in Fig.~\ref{Shape}, illustrating
their counterclockwise winding direction in chart $(x, y')$
according to Eq.~(\ref{eq-modifiedtrajectory}) with parameters
${r=1.5,\omega=\pi/10,\phi_{0}=\pi}$. The shapes of the
time-modulated system parameters can be readily observed using
current experimental
techniques~\cite{experiment-realize1,experiment-realize2,experiment-realize3},
as they exhibit a remarkable simplicity.

\begin{figure}
\scalebox{0.44}{\includegraphics{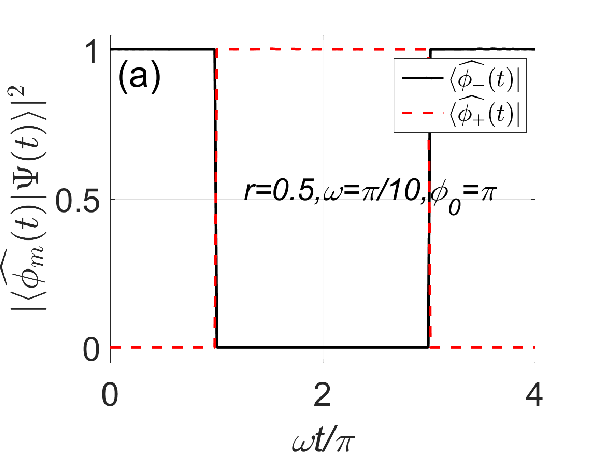}\includegraphics{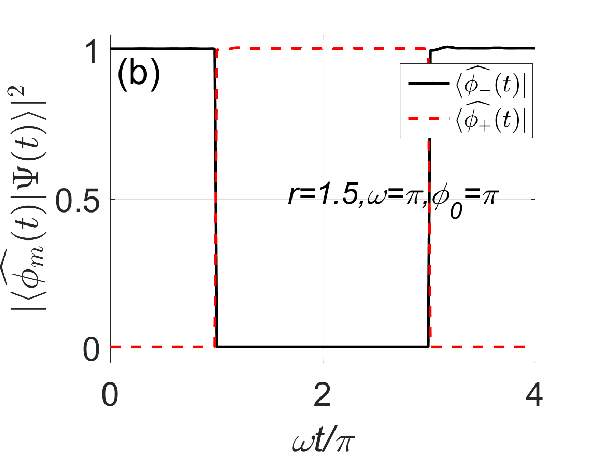}}
\scalebox{0.44}{\includegraphics{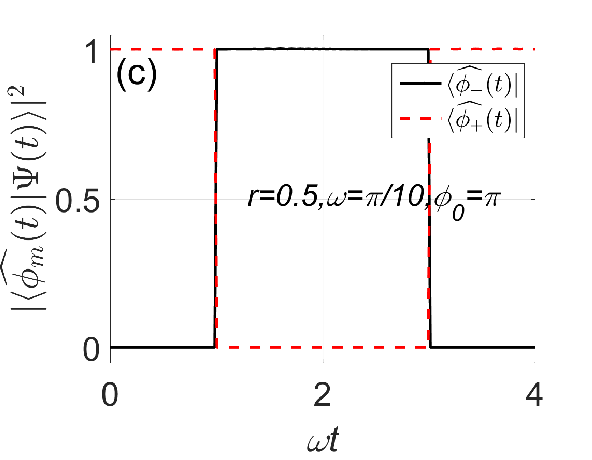}\includegraphics{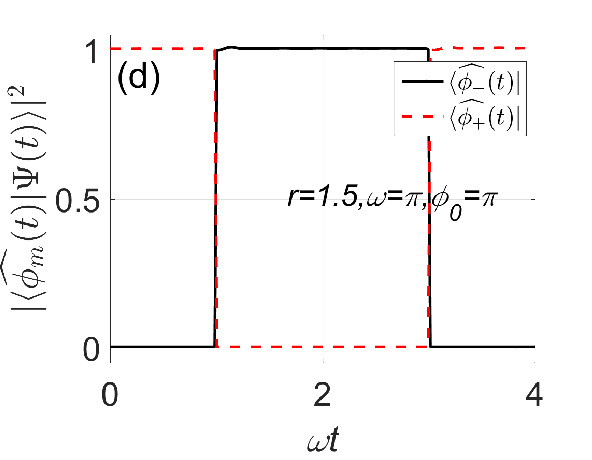}}
\caption{\label{fig-Hm-statetransfer}   The time evolution of the
fidelity $F_{n}=|\langle\widehat{{\phi_{n}}}(t)|\Psi(t)\rangle|^2$
for the time-depend right eigenstate $|\phi_{n}(t)\rangle$
$(n=+,-)$. The initialized state in (a) and (b) [(c) and (d)] is
chosen as $|\Psi(0)\rangle$=$|\phi_{-}(0)\rangle$
[$|\Psi(0)\rangle$=$|\phi_{+}(0)\rangle$], the parameters in (a)
and (c) [(b) and (d)] are chosen as
$\{r=0.5,\omega=\pi/10,\phi_{0}=\pi\}$
$[\{r=1.5,\omega=\pi,\phi_{0}=\pi\}]$, respectively. }
\end{figure}

In Fig.~\ref{fig-Hm-statetransfer}, we present the temporal
evolution of the fidelity for the adiabatic states with two sets
of parameters $({r=0.5,\omega=\pi/10,\phi_{0}=\pi})$ and
$({r=1.5,\omega=\pi,\phi_{0}=\pi})$ under different initial
states. It is evident that, in contrast to the conventional
adiabatic state transfer case (see
Fig.~\ref{fig-adiabaticstatetransfer}), which fails when either
$\omega$ or $r$ takes large values, a complete adiabatic transfer,
i.e., $|\phi_{-}(t)\rangle\leftrightarrow |\phi_{+}(t)\rangle$,
can always be achieved after one period $T=2\pi/\omega$,
regardless of the initialized adiabatic states and control
parameters employed, thus validating our theoretical deduction.
Furthermore, it is noteworthy that the initial state eventually
returns to itself after two periods ($2T=4\pi/\omega$).

\begin{figure}
\scalebox{0.22}{\includegraphics{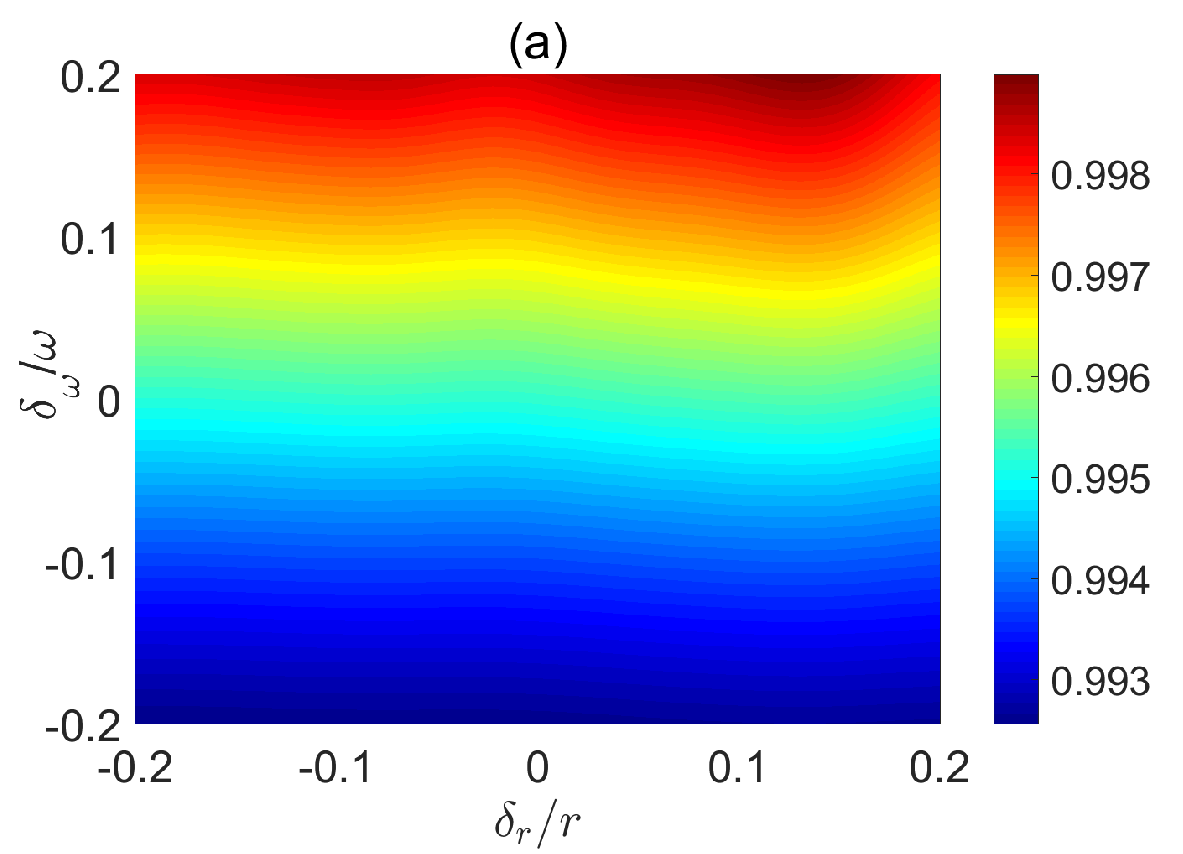}\includegraphics{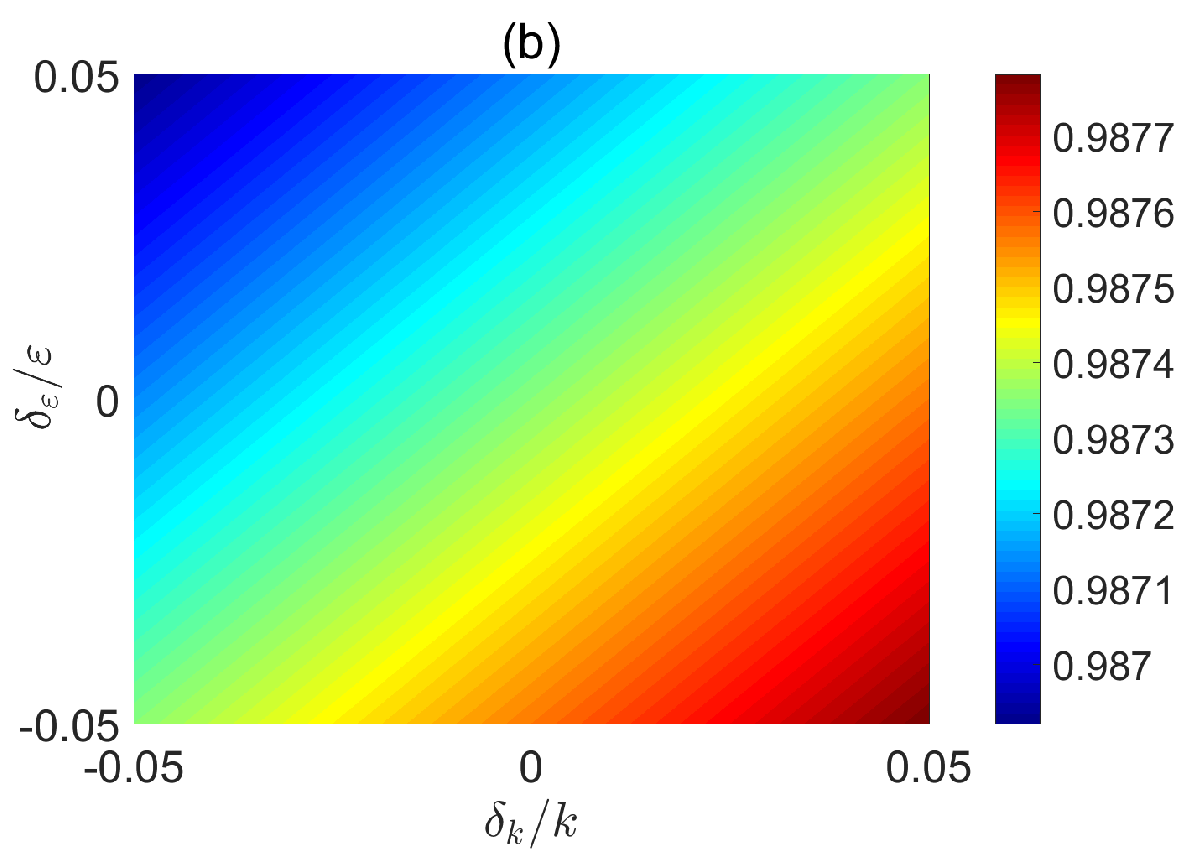}}
\scalebox{0.22}{\includegraphics{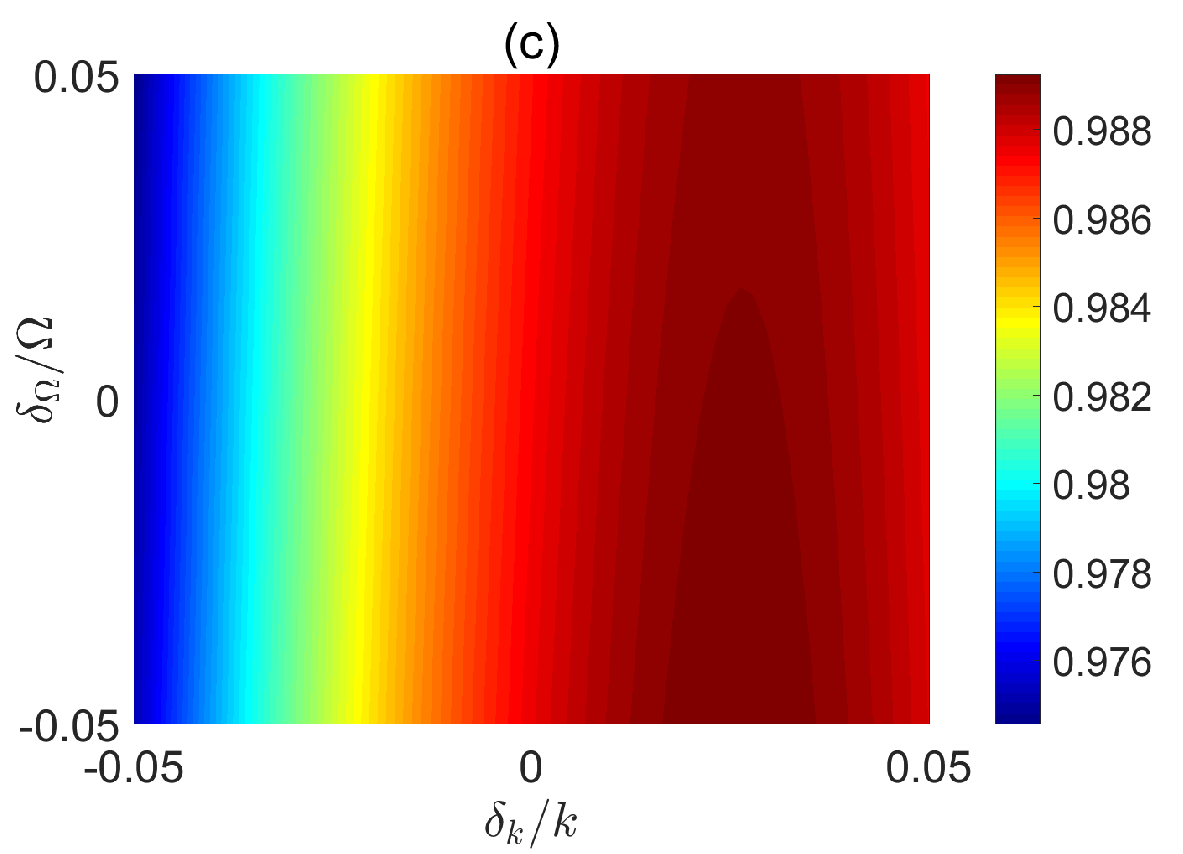}\includegraphics{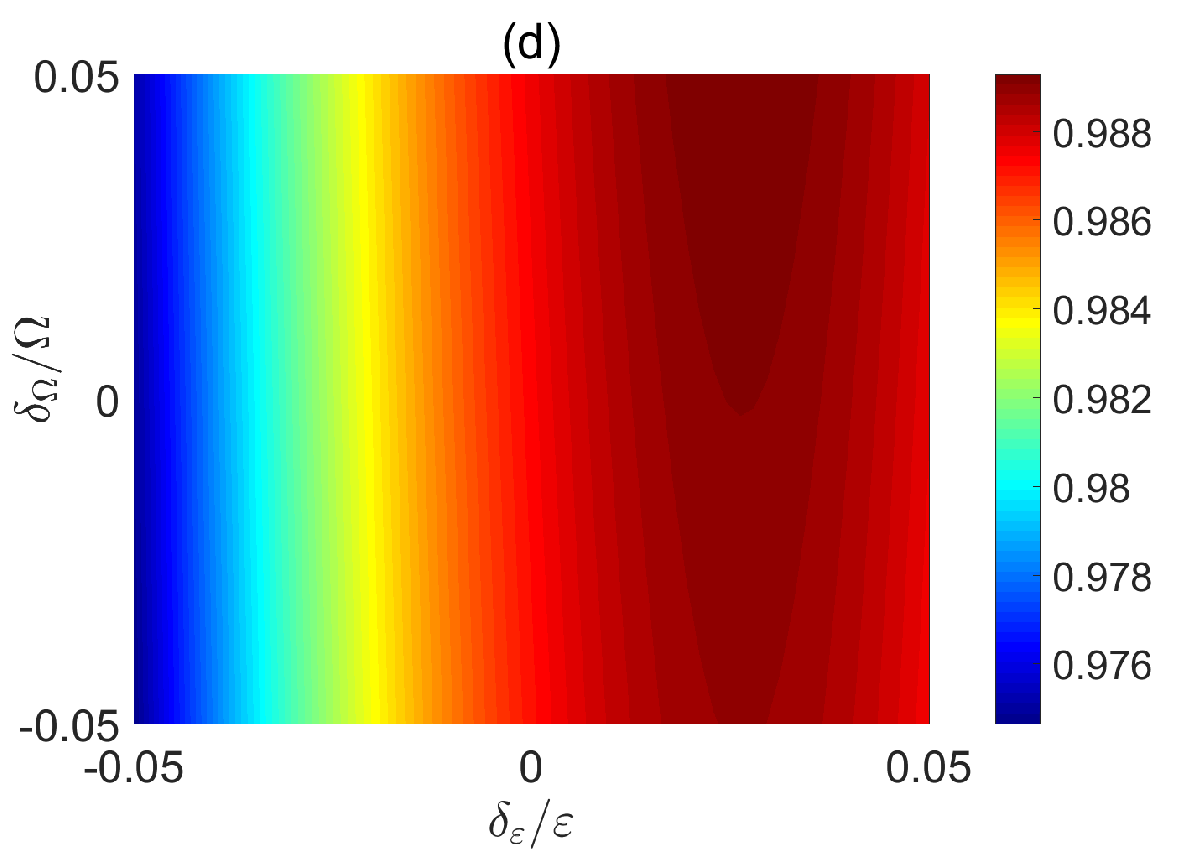}}
\caption{\label{Fsensitivity}   The fidelity of
$F_{+}=|\langle\widehat{{\phi_{+}}}(T)|\Psi(T)\rangle|^2$ vs
several relative deviations  in the control parameters.
(a)$\delta_{r}/r$, $\delta_{\omega}/\omega$, (b)$\delta_{k}/k$,
$\delta_{\varepsilon}/\varepsilon$, (c)$\delta_{k}/k$,
$\delta_{\Omega}/\Omega$, (d)$\delta_{\varepsilon}/\varepsilon$,
$\delta_{\Omega}/\Omega$. The other parameters  are chosen as
${r=1.5, \omega=100\pi, \phi_{0}=\pi}$. }
\end{figure}

In the preceding discussion, the potential impact of parameter
fluctuations has not been thoroughly examined. Therefore, it is
imperative to investigate the sensitivity of the shortcuts to
adiabatic state transfer  in relation to simulated variations in
control parameters. Initially, we focus on analyzing the
fluctuations in trajectory parameters $\delta_{r}/r$ and
$\delta_{\omega}/\omega$. Figure~\ref{Fsensitivity}(a)
demonstrates that the fidelity
$F_{+}$=$|\langle\widehat{{\phi_{+}}}(T)|\Psi(T)\rangle|^2$
remains largely unaffected by these trajectory parameter
fluctuations, with only minor variations at an order of magnitude
around $10^{-3}$. This observation suggests that our designed
system time-evolution trajectory is highly efficient and
guarantees a complete eigenstate transfer even when there exists a
range for trajectory parameters. Furthermore, we also consider
fluctuations in control parameters as described by
Eq.~(\ref{eq-Hm}). The results depicted in
Figs.~\ref{Fsensitivity}(a)-(c) reveal that a high fidelity (above
0.984) can be achieved within a broad range of control parameters
indicated by the red area. Consequently, this scheme exhibits
remarkable efficiency while relaxing constraints on control
parameters.

\section{CONCLUSION}\label{section:IV}

We have proposed a scheme to achieve robustness and fast adiabatic
state transfer in time-modulated two-level non-Hermitian systems,
by appropriately designing the system Hamiltonian and
time-evolution trajectory, and considering the effects of
non-adiabatic transitions. This approach contrasts with
conventional encircling schemes, where the complex system
spectrum, rapid winding speed, and vanishing eigenvalue
differences (at exceptional points) often lead to significant
nontrivial non-adiabatic transitions during the state evolution.
Inspired by Ref.~\cite{Arkhipov-encircling2}, we have transformed
the NH Hamiltonian into a pseudo-Hermitian one with a real
spectrum by suitably mapping the system parameter space onto a
four-dimensional hyperboloid manifold.  We have showed that
potential nonadiabatic couplings under fast winding speeds can be
nullified through a designed coherent control based on STA.
Furthermore, we have designed the system's time-evolution
trajectory to avoid crossing exceptional points. Remarkably, these
procedures enable complete adiabatic state transfers even in
nonadiabatic scenarios after just one period for different
initialized adiabatic states. The effects of variations in control
parameters have been also extensively discussed. Our results
demonstrate that this scheme exhibits insensitivity to moderate
fluctuations of control parameters and one can obtain high
fidelity over a wide range of parameter values. Therefore, the
scheme is powerful and reliable  for quantum-state engineering in
non-Hermitian systems.

\section*{ACKNOWLEDGEMENT}
This work was supported by National Natural Science Foundation of
China (NSFC) (Grants Nos. 12264040, 12374333, 12474366, 12204311
and U21A20436), Jiangxi Natural Science Foundation (Grant Nos.
20232BCJ23022, 20224BAB201027, 20224BAB211025 and 20212BAB211018),
Innovation Program for Quantum Science and Technology (Grant
No.~2021ZD0301200) and the Jiangxi Province Key Laboratory of
Applied Optical Technology (Grant No.~2024SSY03051).


\begin{thebibliography}{999}

\bibitem{NH1}T. Kato, \textit{Perturbation Theory for Linear Operators}, Classics in Mathematics (Springer, Berlin, 1995).
\bibitem{NH2}Y. Ashida, Z. Gong, and M. Ueda, Non-Hermitian physics, Adv. Phys. \textbf{69}, 249 (2020).
\bibitem{NH3} Q. C. Wu,  J. L. Zhao,  Y. L. Fang,  Y. Zhang,  D. X. Chen, C. P. Yang and F. Nori, Extension of Noether's theorem in PT-symmetry systems and its experimental demonstration in an optical setup, Sci. China-Phys. Mech. Astron. \textbf{66}(4), 240312 (2023).

\bibitem{EPs1} L. Feng,  R. El-Ganainy and  L. Ge, Non-Hermitian photonics based on parity-time symmetry, Nat. Photon. \textbf{11}, 752 (2017).
\bibitem{EPs2}$\c{S}$. \"{O}zdemir,  S. Rotter, F. Nori, and L. Yang, Parity-time symmetry and exceptional points in photonics, Nat. Mater. \textbf{18}, 783 (2019).
\bibitem{EPs3}W. D. Heiss, Repulsion of resonance states and exceptional points, Phys. Rev. E \textbf{61}, 929 (2000).
\bibitem{EPs4}H. Cartarius, J. Main, and G. Wunner, Exceptional points in atomic spectra, Phys. Rev. Lett. \textbf{99}, 173003 (2007).

\bibitem{vibrational-cooling}O. Atabek, R. Lefebvre, M. Lepers, A. Jaouadi, O. Dulieu, and V. Kokoouline, Proposal for a laser control of vibrational cooling in Na$_{2}$ using resonance coalescence, Phys. Rev. Lett. \textbf{106}, 173002 (2011).
\bibitem{sensitivity}H. Hodaei, A. U. Hassan, S. Wittek, H. Garcia-Gracia, R. El-Ganainy, D. N. Christodoulides, and M. Khajavikhan, Enhanced sensitivity at higher-order exceptional points, Nature \textbf{548}, 187 (2017).
\bibitem{invisibility}Z. Lin, H. Ramezani, T. Eichelkraut, T. Kottos, H. Cao, and D. N. Christodoulides, Unidirectional invisibility induced by PT-symmetric periodic structures, Phys. Rev. Lett. \textbf{106}, 213901 (2011).

\bibitem{lossinduced}A. Guo, G. J. Salamo, D. Duchesne, R. Morandotti, M. Volatier-Ravat, V. Aimez, G. A. Siviloglou, and D. N. Christodoulides, Observation of PT-symmetry breaking in complex optical potentials, Phys. Rev. Lett. \textbf{103}, 093902 (2009).

\bibitem{topology1} E. J. Bergholtz, J. C. Budich, and F. K. Kunst, Exceptional topology of non-Hermitian systems, Rev. Mod. Phys. \textbf{93}, 015005 (2021).
\bibitem{topology-encircling}C. Guria, Q. Zhong,  $\c{S}$. K. \"{O}zdemir, Y. S. S. Patil, R. El-Ganainy, and J. G. Emmet Harris, Resolving the topology of encircling multiple exceptional points, Nat. Commun. \textbf{15}, 1369 (2024).

\bibitem{Berry-encircling}M. V. Berry, Optical polarization evolution near a nonHermitian degeneracy, J. Opt. A \textbf{13}, 115701 (2011).
\bibitem{Xu-encircling}H. Xu, D. Mason, Luyao Jiang, and J. G. E. Harris, Topological energy transfer in an optomechanical system with exceptional points, Nature (London) \textbf{537}, 80 (2016).
\bibitem{Li-encircling} A. Li,  J. Dong, J. Wang,  Z. Cheng,  J. S. Ho,  D. Zhang, et al., Hamiltonian hopping for efficient chiral mode switching in encircling exceptional points, Phys. Rev. Lett. \textbf{125}(18), 187403 (2020).
\bibitem{Feilhauer-encircling}J. Feilhauer, A. Schumer, J. Doppler, A. A. Mailybaev, J. B\"{o}hm, U. Kuhl,  N. Moiseyev, and S. Rotter, Encircling exceptional points as a non-Hermitian extension of rapid adiabatic passage, Phys. Rev. A \textbf{102}, 040201 (2020).
\bibitem{Arkhipov-encircling1} I. I. Arkhipov, A. Miranowicz, F. Minganti, $\c{S}$. \"{O}zdemir, and F. Nori, Dynamically crossing diabolic points while encircling exceptional curves: A programmable symmetricasymmetric multimode switch, Nat. Commun. \textbf{14}, 2076 (2023).
\bibitem{Ergoktas-encircling} M. S. Ergoktas, S. Soleymani, N. Kakenov, K. Wang, T. B. Smith, G. Bakan, S. Balci, A. Principi, K. S. Novoselov, S. K. Ozdemir, and C. Kocabas, Topological engineering of terahertz light using electrically tunable exceptional point singularities, Science \textbf{376}, 184 (2022).

\bibitem{Tang-encircling}Z. Tang, T. Chen, and X. Zhang, Highly efficient transfer of quantum state and robust generation of entanglement state around exceptional lines, Laser Photonics Rev. 2300794 (2023).
\bibitem{Hassan-encircling}A. U. Hassan, G. L. Galmiche, G. Harari and P. LiKamWa, Chiral state conversion without encircling an exceptional point, Phys. Rev. A \textbf{96}, 052129 (2017).
\bibitem{Hassan-encircling2}A. U. Hassan, B. Zhen, M. Solja\v{c}i\'{c}, M. Khajavikhan, and D. N. Christodoulides, Dynamically encircling exceptional points: Exact evolution and polarization state conversion, Phys. Rev. Lett. \textbf{118}, 093002 (2017).
\bibitem{Zhang-encircling} X. L. Zhang, T. Jiang, and C. T. Chan, Dynamically encircling an exceptional point in anti-parity-time symmetric systems: Asymmetric mode switching for symmetrybroken modes, Light Sci. Appl. \textbf{8}, 88 (2019).
\bibitem{Nasari-encircling}H. Nasari, G. L. Galmiche, H. E. L. Aviles, A. Schumer, A. U. Hassan, Q. Zhong, S. Rotter, P. LiKamWa, D. N. Christodoulides and M. Khajavikhan, Observation of chiral state transfer without encircling an exceptional point, Nature \textbf{605}, 256 (2022).
\bibitem{Arkhipov-encircling2} I. I. Arkhipov, A. Miranowicz, F. Minganti, $\c{S}$. \"{O}zdemir, and F. Nori, Restoring adiabatic state transfer in time-modulated non-hermitian systems, Phys. Rev. Lett. \textbf{133}, 113802, (2024)

\bibitem{open1} F. Verstraete,  M. M. Wolf and  J. I. Cirac, Quantum computation and quantum-state engineering driven by dissipation, Nat. Phys. \textbf{5}(9), 633 (2009).
\bibitem{open2}Q. C. Wu, Y. H. Zhou, B. L. Ye, T. Liu and C. P. Yang, Nonadiabatic quantum state engineering by time-dependent decoherence-free subspaces in open quantum systems, New J. Phys. \textbf{23}, 113005 (2021).
\bibitem{open3}H. Zhang, X. K. Song, Q. Ai, H. Wang, G. J. Yang, and F. G. Deng, Fast and robust quantum control for multimode interactions using shortcuts to adiabaticity, Opt. Express \textbf{27}, 7384 (2019).

\bibitem{criterion1} S. Ib\'{a}\~{n}ez and J. G. Muga, Adiabaticity condition for non-Hermitian Hamiltonians, Phys. Rev. A \textbf{89}(3), 033403 (2014).
\bibitem{criterion2} Q. C. Wu, Y. H. Chen, B. H. Huang, Y. Xia, and J. Song, Reverse engineering of a nonlossy adiabatic Hamiltonian for non-Hermitian systems, Phys. Rev. A \textbf{94}, 053421 (2016).
\bibitem{criterion3}H. Li, H. Z. Shen, S. L. Wu, and X. X. Yi, Shortcuts to adiabaticity in non-Hermitian quantum systems without rotating-wave approximation, Opt. Express \textbf{25}, 30135 (2017).
\bibitem{criterion4}Y. H. Chen, Q. C. Wu, B. H. Huang, J. Song, Y. Xia, and S. B. Zheng, Improving shortcuts to non-Hermitian adiabaticity for fast population transfer in open quantum systems, Ann. Phys. (Berlin) \textbf{530}, 1700247 (2017).
\bibitem{criterion5}Luan, T. Z., H. Z. Shen, and X. X. Yi, Shortcuts to adiabaticity with general two-level non-Hermitian systems, Phys. Rev. A, \textbf{105}, 013714 (2022).

\bibitem{STA1}X. Chen, I. Lizuain, A. Ruschhaupt, D. Gu\'{e}ry-Odelin, and J. G. Muga, Shortcut to adiabatic passage in twoand three-level atoms, Phys. Rev. Lett. \textbf{105}, 123003 (2010).
\bibitem{STA2} E. Torrontegui, S. Ib\'{a}\~{n}ez, S. Mart\'{i}nez-Garaot, M. Modugno, A. del campo, D. Gu\'{e}ry-Odelin, A. Ruschhaupt, X. Chen, and J. G. Muga, Shortcuts to adiabaticity, Adv. At. Mol. Opt. Phys. \textbf{62}, 117 (2013).
\bibitem{STA3} D. Gu\'{e}ry-Odelin,  A. Ruschhaupt,  A. Kiely,  E. Torrontegui,  S. Mart\'{i}nez-Garaot and J. G.  Muga,  Shortcuts to adiabaticity: Concepts, methods, and applications,  Rev. Mod. Phys. \textbf{91}(4) 045001 (2019).

\bibitem{TQD1}M. Demirplak and S. A. Rice, Adiabatic population transfer with control fields, J. Phys. Chem. A \textbf{107}, 993715 (2003).
\bibitem{TQD2} M. V. Berry,  Transitionless quantum driving, J. Phys. A \textbf{42}(36), 365303 (2009).
\bibitem{TQD3}G. Q. Li, G. D. Chen, P. Peng, and W. Qi, NonHermitian shortcut to adiabaticity of two-and threelevel systems with gain and loss, Eur. Phys. J. D \textbf{71}, 14 (2017).
\bibitem{TQD4}Y. H. Chen,  W. Qin,  X. Wang,  A. Miranowicz and  F. Nori,  Shortcuts to adiabaticity for the quantum rabi model: efficient generation of giant entangled cat states via parametric amplification, Phys. Rev. Lett. \textbf{126}(2) 023602 (2021).

\bibitem{LR1}H. R. Lewis and W. B. Riesenfeld, An exact quantum theory of the time-dependent harmonic oscillator and of a charged particle in a time-dependent electromagnetic field, J. Math. Phys. \textbf{10}, 1458 (1969).
\bibitem{LR2}X. Chen,  E. Torrontegui and  J. G. Muga, Lewis-Riesenfeld invariants and transitionless quantum driving, Phys. Rev. A \textbf{83}(6),n062116 (2011).
\bibitem{LR3}Y. H. Kang, Y. H. Chen, B. H. Huang, J. Song, and Y. Xia, Invariant-based pulse design for three-level systems without the rotating-wave approximation, Ann. Phys. (Berlin) \textbf{529}, 1700004 (2017).

\bibitem{Fast1} S. Masuda and K. Nakamura, Fast-forward problem in quantum mechanics, Phys. Rev. A \textbf{78}, 062108 (2008).
\bibitem{Fast2} J. J. Zhu and X. Chen, Fast-forward scaling of atommolecule conversion in Bose-Einstein condensates, Phys. Rev. A \textbf{103}, 023307 (2021).

\bibitem{experiment-realize1}W. Chen, $\c{S}$. \"{O}zdemir, G. Zhao, J. Wiersig, and L. Yang, Exceptional points enhance sensing in an optical microcavity, Nature (London) \textbf{548}, 192 (2017).
\bibitem{experiment-realize2} I. I. Arkhipov, A. Miranowicz, F. Nori, $\c{S}$. K. \"{O}zdemir, and F. Minganti, Fully solvable finite simplex lattices with open boundaries in arbitrary dimensions, Phys. Rev. Res. \textbf{5}, 043092 (2023).
\bibitem{experiment-realize3}H. S. Xu and L. Jin, Coupling-induced nonunitary and unitary scattering in anti-PT-symmetric non-Hermitian systems, Phys. Rev. A \textbf{104}, 012218 (2021).

\bibitem{map1} M. V. Berry, Physics of non-Hermitian degeneracies, Czech. J. Phys. \textbf{54}, 1039 (2004).
\bibitem{map2} J. Maldacena, The large-N limit of superconformal field theories and supergravity, Adv. Theor. Math. Phys. \textbf{2}, 231 (1998).

\bibitem{Biorthogonal1} D. C. Brody, Biorthogonal quantum mechanics, J. Phys. A-Math. Theor. \textbf{47}, 035305 (2013).


\end{thebibliography}
\end{document}